\documentclass[traditabstract,onecolumn]{aa}

\begin{document}

\title{Solar wind and dynamics of meteoroid}
\author{J.~Kla\v{c}ka}
\institute{Department of Astronomy Physics of the Earth, and Meteorology, \\
   Faculty of Mathematics, Physics and Informatics, Comenius University, \\
   Mlynsk\'{a} dolina, 842~48 Bratislava, Slovak Republic \\
   e-mail: klacka@fmph.uniba.sk }

\date{}

\abstract
{
The effect of solar wind on dust particle, meteoroid, is investigated.
Rotation of the particle is also considered. Detail derivations of both
equation of motion and Euler's dynamical equations, are presented.
The most simple form of the action of the solar wind is considered and
the derived equation of motion reduces to ``pseudo-Poynting-Robertson effect''
for non-rotating spherical particle.
The final equation of motion contains also angular velocity term and it may
cause change of orbital plane of the meteoroid. Analysis of the Euler's
dynamical equations shows that rotational axis of the meteoroid changes in time
and equilibrium state for the orientation of the axis of rotation does not
exist for spherical particle.
}


\authorrunning{J.~Kla\v{c}ka}
\titlerunning{Solar wind and dynamics of meteoroid}
\maketitle

\section{Introduction}
At the beginning of the past century, Poynting (1903) has formulated
a problem of finding equation of
motion for a perfectly absorbing spherical particle under the action
of electromagnetic radiation. Robertson (1937) derived
a correct equation of motion for a perfectly absorbing spherical
particle. This result has been applied to astronomical problems
for several decades and is known as the Poynting-Robertson (P-R)
effect. Relativistically covariant equation of motion for
general properties of spherical particles was presented by Kla\v{c}ka
(2000, 2004). Complete understanding of the action of the incoming
electromagnetic radiation on the motion of the spherical dust particle was
obtained by Kla\v{c}ka (2008a, 2008b) and Kla\v{c}ka {\it et al.} (2009a).
Partly similar effect on interplanetary dust particles
is caused not only by solar electromagnetic radiation, but also by stream of
charged particles flowing from the Sun. This stream continuously emanating
from the Sun's surface consists mostly from electrons and protons, $\alpha-$
particles and heavier atoms. The stream is called also corpuscular solar radiation,
or, shortly, solar wind. Dynamical effects of solar wind on
motion of meteoroids are similar to dynamical effects of solar electromagnetic
radiation and they are known also as ``pseudo-Poynting-Robertson effect''
(Dohnanyi 1978). Detail understanding of the effect of solar wind
on motion of spherical interplanetary dust particle is presented in
Kla\v{c}ka {\it et al.} (2009b).

All the mentioned equations of motion for dust particles under action of
electromagnetic radiation and solar wind are correct for non-rotating dust
grains. What is the effect of rotation? How does dust particle rotate?
In order to be able to answer the questions, we will deal with the effect of
rotation on the dust particle. We will obtain results not only for time
evolution of rotation of the particle, but also equation of motion of the
rotating particle.

\section{Proper reference frame of the dust particle -- stationary particle}
We will use considerations and accuracy to the first
order in $\vec{v} / u$ and $\vec{\omega} R / u$ ($\vec{v}$ is
heliocentric velocity of the spherical dust particle of radius $R$, $u$ is
the speed of solar wind with respect to the Sun and $\vec{\omega}$ is
angular velocity of the rotation of the particle about a rotational axis
containing center of mass of the particle).

The term ``stationary particle'' will denote a particle which does not
exhibit translational motion in a given inertial frame of reference.
Primed quantities will denote quantities measured in the
proper reference frame of the center of mass of the particle.
The particle may be characterized with a rotational motion
in the proper reference frame of the center of mass.

Let $S'$ is an incident flux density of solar wind energy
(energy flow through unit area perpendicular to the ray per unit time)
and geometrical cross section of the spherical particle is $\pi$ $R^{2}$.
We have $S'$ $=$ $n'$ $m_{1}$ $| \vec{u}' |^{2}$ $c$, where $n'$ is
concentration (number of particles per unit volume) of solar wind particles
of mass $m_{1}$ (for simplicity of these considerations only one type of
particles is considered), $c$ is the speed of light,
$\vec{u}'$ $=$ $u'$ $\vec{S}'$ is
velocity vector of solar wind particles:
unit vector $\vec{S}'$ denotes direction and orientation of the incident
beam of solar wind particles measured in the proper reference frame of the
center of mass of the dust particle. Relations (1) -- (19) in Kla\v{c}ka and
Saniga (1993) are significant for the effect of incident solar wind
particle. Moreover, Lorentz transformation yields for elementary force 
acting on the elementary area of the spherical particle due to action 
of the incident solar wind $\left ( d \vec{p}' / d t \right ) _{el;inc}$ $=$
$\left ( d \vec{p}'' / d t \right ) _{el;inc}$ $+$
$E''_{inc} \vec{v}'_{rot;el} / c^{2}$, where double primed quantities are
measured in the frame of reference connected with the rotating
(rotational velocity $\vec{v}'_{rot;el}$) elementary area on the sphere.
We have
$\left ( d \vec{p}'' / d t \right ) _{el;inc}$ $=$ $n'_{el}$
$[ ( d A' )_{el} \cos \alpha']$ $| \vec{u}_{el}' |$ $m_{1}$ $\vec{u}_{el}'$,
$n_{el}'$ $=$ $n'$ $=$ $n$ (within the considered accuracy),
$\vec{u}_{el}'$ $=$ $\vec{u}'$ $-$ $\vec{\omega}$ $\times$ $\vec{\xi} '$.
If we take all these results into account, then we can write
for an elementary force caused by the incident solar wind
\begin{eqnarray}\label{1}
\left ( \frac{d~ \vec{p}'}{d~t} \right ) _{el;inc} &=&
      n' ~m_{1} ~ \vec{u}'^{2}
      \left [ \left ( d A' \right )_{el} \cos \alpha' \right ] \times
\nonumber \\
& & \times \left \{ \vec{S}' ~-~
    \left ( \frac{\vec{\omega} \times \vec{\xi} '}{| \vec{u}' |} \cdot \vec{S}'
    \right ) \vec{S}' ~-~
    \frac{\vec{\omega} \times \vec{\xi} '}{| \vec{u}' |} \right \} ~+~
\nonumber \\
& & +~ n' ~m_{1} ~ | \vec{u}' | ~
      \left [ \left ( d A' \right )_{el} \cos \alpha ' \right ]
    ~ \vec{\omega} \times \vec{\xi} ' ~,
\end{eqnarray}
where $\vec{\xi} '$ is position vector of the element on the surface of the spherical
particle with respect to the center of mass of the particle,
$\left ( d A' \right )_{el}$ is area of the element. Within the considered
accuracy, Eq. (1) is equivalent to
\begin{equation}\label{2}
\left ( \frac{d~ \vec{p}'}{d~t} \right ) _{el;inc} =
      \frac{S' \left ( d A' \right )_{el} \cos \alpha '}{c} ~ \left ( 1 ~-~
      \frac{\vec{\omega} \times \vec{\xi} '}{| \vec{u}' |}
      \cdot \vec{S}' \right ) ~ \vec{S}' ~,
\end{equation}
if the incident flux density of solar wind energy
$S'$ $=$ $n'$ $m_{1}$ $| \vec{u}' |^{2}$ $c$ is used: this enable to consider
more general composition of solar wind, as for various types of solar wind
particles.

Eq. (2) enables to find the total force of the incident solar wind
on the dust particle, and, also the total torque acting on the particle:
\begin{equation}\label{3}
\left ( \frac{d~ \vec{p}'}{d~t} \right ) _{total;inc} =
\sum_{el} \left ( \frac{d~ \vec{p}'}{d~t} \right ) _{el;inc} ~,
\end{equation}
\begin{equation}\label{4}
\vec{M}'_{total;inc} =
\sum_{el} \vec{\xi} ' \times \left ( \frac{d~ \vec{p}'}{d~t} \right ) _{el;inc} ~.
\end{equation}

In order to find $\left ( d \vec{p}' / d t \right ) _{total;inc}$ and
$\vec{M}'_{total;inc}$, we may perform calculations in some special frame
of reference:
\begin{eqnarray}\label{5}
\left ( d A' \right )_{el} &=&
   \left [ \left ( R~ \sin \alpha ' \right )~ d~ \lambda ' \right ] ~ R~ d~\alpha ' ~,
\nonumber \\
\vec{\xi} ' &=& R ~ \left ( \cos \alpha ' , ~ \cos \lambda ' ~ \sin \alpha ' ,~
		\sin \lambda ' ~ \sin \alpha ' \right ) ~,
\nonumber \\
\vec{S}' &=& -~ \left ( 1, ~0, ~0 \right ) ~,
\nonumber \\
\vec{\omega} &=& \left ( \omega_{x} ,~\omega_{y} ,~\omega_{z} \right ) ~,
\nonumber \\
\alpha ' &\in& \langle 0,~ \pi ~/ ~2 \rangle ~,
\nonumber \\
\lambda ' &\in& \langle 0,~ 2~ \pi ) ~.
\end{eqnarray}
Putting Eqs. (2) and (5) into Eqs. (3) and (4), using also
$\sum_{el}$ $\longrightarrow$ $\int$, one finally obtains
\begin{equation}\label{6}
\left ( \frac{d~ \vec{p}'}{d~t} \right ) _{total;inc} =
      \frac{S' ~\pi ~R^{2}}{c} ~\vec{S}' ~,
\end{equation}
\begin{equation}\label{7}
\vec{M}' _{total;inc} = -~ \frac{1}{4} ~
    \frac{S' ~\pi ~R^{4}}{c~| \vec{u}' |} ~\left \{ \vec{\omega} ~-~
    \left ( \vec{\omega} \cdot \vec{S} ' \right ) ~ \vec{S} ' \right \} ~.
\end{equation}

As for the ``outgoing part'' of solar wind, we will suppose that the solar wind
does not impart any momentum to the corresponding element of the area on the
surface of the dust particle. Thus, we have
$\left ( d \vec{p}'' / d t \right ) _{el;out}$ $=$ 0
for elementary force acting on the elementary area of the spherical
particle due to action of the ``outgoing'' solar wind.
Using the relation $\left ( d \vec{p}' / d t \right ) _{el;out}$ $=$
$\left ( d \vec{p}'' / d t \right ) _{el;out}$ $+$
$E''_{out} \vec{v}'_{rot;el} / c^{2}$, we can write
\begin{equation}\label{8}
\left ( \frac{d~ \vec{p}'}{d~t} \right ) _{el;out} =
      \frac{x'~ S'}{c}  ~
      \left [ \left ( d A' \right )_{el} \cos \alpha' \right ]
      \frac{\vec{\omega} \times \vec{\xi} '}{| \vec{u}' |} ~,
\end{equation}
where it is supposed that the $x'-th$ part of the incident energy per unit time
is lost from the dust particle (see Eq. 20 in Kla\v{c}ka and Saniga 1993).

Eq. (8) enables to find the total force acting on the particle due to the
``outgoing'' solar wind, and, also the total torque acting on the particle (the
sign minus is due to the conservation of momentum):
\begin{equation}\label{9}
\left ( \frac{d~ \vec{p}'}{d~t} \right ) _{total;out} =
\sum_{el} \left \{ ~-~ \left ( \frac{d~ \vec{p}'}{d~t} \right ) _{el;out}
	  \right \} ~,
\end{equation}
\begin{equation}\label{10}
\vec{M}'_{total;out} =
\sum_{el} \vec{\xi} ' \times  \left \{ ~-~
\left ( \frac{d~ \vec{p}'}{d~t} \right ) _{el;out} \right \} ~.
\end{equation}

Detail calculations in a special frame of reference, e. g. given by
relations (5), yield
\begin{equation}\label{11}
\left ( \frac{d~ \vec{p}'}{d~t} \right ) _{total;out} =
      \frac{x' ~ S' ~ \pi ~ R^{2}}{c} ~\frac{2}{3} ~\frac{R}{| \vec{u}' |} ~
      \vec{\omega} \times \vec{S} ' ~,
\end{equation}
\begin{equation}\label{12}
\vec{M}' _{total;out} = -~ \frac{1}{4} ~
    \frac{x'~ S' ~\pi ~R^{4}}{c ~| \vec{u}' |} ~ \left \{
    3~ \vec{\omega} ~-~
    \left ( \vec{\omega} \cdot \vec{S} ' \right ) ~ \vec{S} ' \right \} ~.
\end{equation}

The total force and torque acting on the particle due to solar wind:
\begin{equation}\label{13}
\left ( \frac{d~ \vec{p}'}{d~t} \right ) _{total} =
\left ( \frac{d~ \vec{p}'}{d~t} \right ) _{total;inc} ~+~
\left ( \frac{d~ \vec{p}'}{d~t} \right ) _{total;out} ~,
\end{equation}
\begin{equation}\label{14}
\vec{M}' _{total} =
\vec{M}' _{total;inc} ~+~
\vec{M}' _{total;out} ~,
\end{equation}
or, using Eqs. (6)-(7) and (11)-(12):
\begin{equation}\label{15}
\left ( \frac{d~ \vec{p}'}{d~t} \right ) _{total} =
      \frac{S' ~\pi ~R^{2}}{c} ~ \left \{ \vec{S}' ~+~
      x' ~\frac{2}{3} ~\frac{R}{| \vec{u}' |} ~
      \vec{\omega} \times \vec{S} ' \right \} ~,
\end{equation}
\begin{equation}\label{16}
\vec{M}' _{total} = -~ \frac{1}{4} ~ \frac{S' ~\pi ~R^{4}}{c ~| \vec{u}' |} ~
\left \{ \left ( 1 ~+~ 3~x' \right ) ~ \vec{\omega} ~-~ \left ( 1 ~+~ x' \right ) ~
\left ( \vec{\omega} \cdot \vec{S} ' \right ) ~ \vec{S} ' \right \} ~.
\end{equation}

\section{Stationary frame of reference -- equation of motion}
By the term ``stationary frame of reference''
(laboratory frame) we mean a frame of reference
in which particle moves with a velocity vector $\vec{v} = \vec{v} (t)$.
The physical quantities measured in the stationary frame of reference
will be denoted by unprimed symbols.

Our aim is to derive equation of motion for the particle in the
stationary frame of reference. We will use the fact that we know
this equation in the proper frame of reference: Eq. (15).
In order to find the equation in the stationary frame of reference,
we will use the Lorentz transformation for force acting on the dust grain:
\begin{equation}\label{17}
\frac{d~ \vec{p}}{d~t} =
			\frac{d~ \vec{p}'}{d~t} ~+~
			\frac{E'_{inc} ~-~E'_{out}}{c} ~\frac{\vec{v}}{c} ~.
\end{equation}
Since
\begin{eqnarray}\label{18}
E'_{out} &=& x' ~ E'_{inc} ~,
\nonumber \\
E'_{inc} &=& n'~ m_{1} ~ | \vec{u}' | ~c^{2} ~ \pi ~ R^{2} ~,
\end{eqnarray}
(compare with the right-hand side of Eq. 1:
$E'_{inc}$ $=$ $\sum_{el} n'~ m_{1} ~ | \vec{u}' | ~c^{2} ~
[ (dA')_{el} \cos \alpha ']$), we have
\begin{equation}\label{19}
\frac{d~ \vec{p}}{d~t} = \frac{d~ \vec{p}'}{d~t} ~+~
   \left ( 1 ~-~ x' \right ) ~ n'~ m_{1} ~ | \vec{u}' | ~ \pi ~ R^{2} ~\vec{v} ~.
\end{equation}
Since $S' = S~ ( 1 ~-~ 2 \vec{v} \cdot \vec{S} /  | \vec{u} | )$,
$\vec{S} ' = ( 1 ~+~ \vec{v} \cdot \vec{S} / | \vec{u} |  )~ \vec{S}$
$-$ $\vec{v} / | \vec{u} |$, Eqs. (15) and (19) yield
\begin{eqnarray}\label{20}
\frac{d~ \vec{p}}{d~t} &=& \frac{S ~\pi ~R^{2}}{c} ~ \left \{
   \left ( 1 ~-~ \frac{\vec{v} \cdot \vec{S}}{| \vec{u} |} \right )~ \vec{S} ~-~
   \frac{\vec{v}}{| \vec{u} |}	~+~ x' ~\frac{2}{3} ~\frac{R}{| \vec{u} |} ~
   \vec{\omega} \times \vec{S} \right \} ~+~
\nonumber \\
& & +~ \left ( 1 ~-~ x' \right ) ~ \frac{S ~\pi ~R^{2}}{c} ~
    \frac{\vec{v}}{| \vec{u} |}  ~.
\end{eqnarray}
The last term in Eq. (20) corresponds to the
change of grain's mass, in accordance with Eq. (18):
\begin{equation}\label{21}
\frac{d~m}{d~t} = \left ( 1 ~-~ x' \right ) ~
		  \frac{S ~\pi ~R^{2}}{c~| \vec{u} |} ~.
\end{equation}
Since $d \vec{p} / d t$ $=$ $d ( m \vec{v} ) / d t$ $=$
$m~ d \vec{v} / d t$ $+$ $\vec{v} ~d m / d t$, Eqs. (20)-(21) lead to
\begin{equation}\label{22}
\frac{d~ \vec{v}}{d~t} = \frac{S ~\pi ~R^{2}}{m~c} ~ \left \{
   \left ( 1 ~-~ \frac{\vec{v} \cdot \vec{S}}{| \vec{u} |} \right )~ \vec{S} ~-~
   \frac{\vec{v}}{| \vec{u} |}	~+~ x' ~\frac{2}{3} ~\frac{R}{| \vec{u} |} ~
   \vec{\omega} \times \vec{S} \right \} ~.
\end{equation}

Mass of the homogeneous spherical particle is $m$ $=$ ( $4 \pi / 3$ )
$\varrho$ $R^{3}$, where $\varrho$ is mass density. Eq. (21) reduces to
\begin{equation}\label{23}
\frac{d~R}{d~t} = -~ \frac{K}{r^{2}} ~,
\end{equation}
where $r$ is distance between the particle and the Sun, and
\begin{equation}\label{24}
K = \left ( x' ~-~ 1 \right ) ~
\frac{r^{2} ~S}{4 ~\varrho~ c~| \vec{u} |}
\end{equation}
is a constant for a given material.
According to Dohnanyi (1978) and Kapi\v{s}insk\'{y} (1984) we
can take $K \approx 4 \times 10^{-9}$ $cm$ $year^{-1}$ (see also Leinert and
Gr\H{u}n 1991). The phenomenon presented in Eq. (23) is known as the
``corpuscular sputtering''. The units used in Eq. (23) are:
[ $R$ ] = $cm$, [ $t$ ] = $year$, [ $r$ ] = $AU$,
[ $K$ ] = $cm$ $year^{-1}$. Eq. (24) enables to find the vale of the
dimensionless factor $x'$: $x'$ $=$ $1.9 \times 10^{11}$
$K [cm~ year^{-1}]$ $\varrho [g~cm^{-3}]$ $+$ 1. As an example,
for the case $K = 4 \times 10^{-9}$ $cm$ $year^{-1}$ and
$\varrho$ $=$ 1 $g$ $cm^{-3}$, we obtain $x'$ $=$ 7.6 $\times 10^{2}$.

\section{Rotational motion of spherical particle}
On the basis of Eq. (16), we can immediately write Euler's dynamical equations:
\begin{equation}\label{25}
\frac{d~ \left [ \left ( 2 / 5 \right ) ~m~R^{2}~ \vec{\omega} \right ]}{d~t}  =
-~ \frac{1}{4} ~ \frac{S ~\pi ~R^{4}}{c ~| \vec{u} |} ~
\left \{ \left ( 1 ~+~ 3~x' \right ) ~ \vec{\omega} ~-~ \left ( 1 ~+~ x' \right ) ~
\left ( \vec{\omega} \cdot \vec{S} ' \right ) ~ \vec{S} ' \right \} ~,
\end{equation}
where we have used $S$ instead of $S'$ and $| \vec{u} |$ istead of $| \vec{u}' |$,
within the considered accuracy. The moment of inertia of a sphere of uniform
density is ($2/5$) $m$ $R^{2}$.

In order to solve Eq. (25), together with Eqs. (22)-(23), we use, as standardly,
Euler's angles $\psi$, $\vartheta$ and $\varphi$. We have
\begin{eqnarray}\label{26}
\frac{d~\omega_{1'}}{d~t}  &=&
    - ~\frac{5 ~\pi}{8} ~\frac{S~R^{2}}{m~c~ | \vec{u} |} ~
\left \{ \left ( 1 + 3 x' \right ) ~ \omega_{1'} ~-~ \left ( 1 + x' \right ) ~
\left ( \vec{\omega} \cdot \vec{S} \right ) ~ S_{1'} \right \}
    ~-~ \frac{5}{R} ~\frac{d~R}{d~t} ~ \omega_{1'} ~,
\nonumber \\
\frac{d~\omega_{2'}}{d~t}  &=&
    - ~\frac{5 ~\pi}{8} ~\frac{S~R^{2}}{m~c~ | \vec{u} |} ~
\left \{ \left ( 1 + 3 x' \right ) ~ \omega_{2'} ~-~ \left ( 1 + x' \right ) ~
\left ( \vec{\omega} \cdot \vec{S} \right ) ~ S_{2'} \right \}
    ~-~ \frac{5}{R} ~\frac{d~R}{d~t} ~ \omega_{2'} ~,
\nonumber \\
\frac{d~\omega_{3'}}{d~t}  &=&
    - ~\frac{5 ~\pi}{8} ~\frac{S~R^{2}}{m~c~ | \vec{u} |} ~
\left \{ \left ( 1 + 3 x' \right ) ~ \omega_{3'} ~-~ \left ( 1 + x' \right ) ~
\left ( \vec{\omega} \cdot \vec{S} \right ) ~ S_{3'} \right \}
    ~-~ \frac{5}{R} ~\frac{d~R}{d~t} ~ \omega_{3'} ~,
\end{eqnarray}
where we have used the fact that scalar product satisfies
$\sum_{p=1}^{3} \omega_{p} S_{p}$ $=$ $\sum_{q=1}^{3} \omega_{q'} S_{q'}$, and,
$\vec{S}$ $=$ $S_{1} \vec{i}$ $+$ $S_{2} \vec{j}$ $+$ $S_{3} \vec{k}$ in the
stationary frame of reference,
$\vec{S}'$ $=$ $S_{1'} \vec{i'}$ $+$ $S_{2'} \vec{j'}$ $+$ $S_{3'} \vec{k'}$ in
the frame of reference rotating with the particle. Moreover, the well-known
relations are:
\begin{eqnarray}\label{27}
\omega_{1'} &=& \dot{\psi} ~ \sin \varphi ~ \sin \vartheta ~+~
		\dot{\vartheta} ~ \cos \varphi ~,
\nonumber \\
\omega_{2'} &=& \dot{\psi} ~ \cos \varphi ~ \sin \vartheta ~-~
		\dot{\vartheta} ~ \sin \varphi ~,
\nonumber \\
\omega_{3'} &=& \dot{\psi} ~ \cos \vartheta ~+~
		\dot{\varphi} ~,
\end{eqnarray}
\begin{eqnarray}\label{28}
\omega_{1} &=& \dot{\varphi} ~ \sin \psi ~ \sin \vartheta ~+~
		\dot{\vartheta} ~ \cos \psi ~,
\nonumber \\
\omega_{2} &=& -~ \dot{\varphi} ~ \cos \psi ~ \sin \vartheta ~+~
		\dot{\vartheta} ~ \sin \psi ~,
\nonumber \\
\omega_{3} &=& \dot{\varphi} ~ \cos \vartheta ~+~
		\dot{\psi} ~,
\end{eqnarray}
\begin{eqnarray}\label{29}
\vec{S} &=& S_{1}~ \vec{i} ~+~ S_{2} ~\vec{j} ~+~S_{3} ~\vec{k} ~,
\nonumber \\
\vec{S} &=& \frac{x}{r}~ \vec{i} ~+~ \frac{y}{r} ~\vec{j} ~+~ \frac{z}{r} ~\vec{k} ~,
\nonumber \\
\vec{r} &=& x~ \vec{i} ~+~ y ~\vec{j} ~+~ z ~\vec{k} ~,
\end{eqnarray}
where $\vec{r}$ is position vector of the particle with respect to the Sun,
source of solar wind, $r = | \vec{r} |$,
$\vec{v} \equiv d \vec{r} / d t$, and,
\begin{equation}\label{30}
 \left ( \begin{array}{c}
S_{1'} \\ S_{2'} \\ S_{3'}
\end{array} \right ) = \left ( \begin{array}{ccc}
		       R_{11} & R_{12} & R_{13} \\
		       R_{21} & R_{22} & R_{23} \\
		       R_{31} & R_{32} & R_{33} \\
		       \end{array} \right )
					      \left ( \begin{array}{c}
					      S_{1} \\ S_{2} \\ S_{3}
					      \end{array} \right ) ~,
\end{equation}
\begin{eqnarray}\label{31}
R_{11} &=& \cos \psi ~ \cos \varphi ~-~ \cos \vartheta ~\sin \psi ~
					\sin \varphi ~,
\nonumber \\
R_{12} &=& \sin \psi ~ \cos \varphi ~+~ \cos \vartheta ~\cos \psi ~
					\sin \varphi ~,
\nonumber \\
R_{13} &=& \sin \vartheta ~\sin \varphi ~,
\nonumber \\
R_{21} &=& -~ \cos \psi ~ \sin \varphi ~-~ \cos \vartheta ~\sin \psi ~
					   \cos \varphi ~,
\nonumber \\
R_{22} &=& -~ \sin \psi ~ \sin \varphi ~+~ \cos \vartheta ~\cos \psi ~
					   \cos \varphi ~,
\nonumber \\
R_{23} &=&  \sin \vartheta ~\cos \varphi ~,
\nonumber \\
R_{31} &=& \sin \vartheta ~\sin \psi ~,
\nonumber \\
R_{32} &=& -~ \sin \vartheta ~\cos \psi ~,
\nonumber \\
R_{33} &=& \cos \vartheta ~.
\end{eqnarray}
Eqs. (28) and (29) yield
\begin{eqnarray}\label{32}
\vec{\omega} \cdot \vec{S} &=& \left ( \dot{\vartheta} ~ \cos \psi ~+~
\dot{\varphi} ~ \sin \vartheta ~ \sin \psi \right ) ~ \frac{x}{r} ~+~
     \left ( \dot{\vartheta} ~ \sin \psi ~-~
     \dot{\varphi} ~ \sin \vartheta ~ \cos \psi \right ) ~ \frac{y}{r} ~+~
\nonumber \\
& & ~+~ \left ( \dot{\varphi} ~ \cos \vartheta ~+~
\dot{\psi}  \right ) ~ \frac{z}{r} ~.
\end{eqnarray}

We can take into account a non-dimensional parameter, for Solar System
\begin{equation}\label{33}
\beta = \frac{r^{2} ~S_{light} ~\pi ~ R^{2}}{G ~M_{\odot} ~m ~c} ~\bar{Q}'_{pr}
      \equiv \frac{L_{\odot} ~\pi ~ R^{2}}{4~ \pi ~G ~M_{\odot} ~m ~c} ~\bar{Q}'_{pr}
	    = 2.4 \times 10^{-3} ~
	    \frac{R^{2} \left [ m^{2} \right ]}{m \left [ kg \right ]} ~\bar{Q}'_{pr} ~,
\end{equation}
where $L_{\odot}$ is the rate of energy outflow
from the Sun, the solar luminosity, $M_{\odot}$ is mass of the Sun and
$G$ is the gravitational constant and $\bar{Q}'_{pr}$ is a dimensionless efficiency
factor for radiation pressure (Kla\v{c}ka 2004).
The non-dimensional parameter (``the ratio of radiation pressure
force to the gravitational force'') reduces to
$\beta = 5.7 \times 10^{-5} ~ \bar{Q}'_{pr} / ( \varrho [g/cm^{3}] ~R [cm] )$,
for homogeneous spherical particle:
$\varrho$ is mass density and $R$ is radius of the sphere.
On the basis of Eq. (33), we can immediately write for the quantities of solar wind:
$S~ \pi~ R^{2} ~/~ ( m~c )$ $=$ $\beta~( \eta ~/~ \bar{Q}'_{pr} )$
($G ~M_{\odot}~/~r^{2}$) $| \vec{u} | / c$, where $\eta \approx$ 1/3.
Inserting this result into Eq. (26),
using also Eq. (24), one immediately obtains
\begin{eqnarray}\label{34}
\frac{d~\omega_{1'}}{d~t}  &=&
 - ~\frac{5}{8~c} ~\beta ~\frac{\eta}{\bar{Q}'_{pr}} ~ \frac{G~M_{\odot}}{r^{2}} ~
\left \{ \frac{x' + 11}{3} ~ \omega_{1'} ~-~ \left ( 1 + x' \right ) ~
\left ( \vec{\omega} \cdot \vec{S} \right ) ~ S_{1'} \right \} ~,
\nonumber \\
\frac{d~\omega_{2'}}{d~t}  &=&
 - ~\frac{5}{8~c} ~\beta ~\frac{\eta}{\bar{Q}'_{pr}} ~ \frac{G~M_{\odot}}{r^{2}} ~
\left \{ \frac{x' + 11}{3}  ~ \omega_{2'} ~-~ \left ( 1 + x' \right ) ~
\left ( \vec{\omega} \cdot \vec{S} \right ) ~ S_{2'} \right \} ~,
\nonumber \\
\frac{d~\omega_{3'}}{d~t}  &=&
 - ~\frac{5}{8~c} ~\beta ~\frac{\eta}{\bar{Q}'_{pr}} ~ \frac{G~M_{\odot}}{r^{2}} ~
\left \{ \frac{x' + 11}{3} ~ \omega_{3'} ~-~ \left ( 1 + x' \right ) ~
\left ( \vec{\omega} \cdot \vec{S} \right ) ~ S_{3'} \right \} ~.
\end{eqnarray}

The orbital and rotational motion of spherical dust particle
under action of solar wind and gravity of the Sun is given by
Eqs. (23), (27)-(34), together with equation
\begin{equation}\label{35}
\frac{d~ \vec{v}}{d~t} =  -~ \frac{G~M_{\odot}}{r^{2}} ~\vec{S} +
   \beta ~\frac{\eta}{\bar{Q}'_{pr}} ~ \frac{G~M_{\odot}}{r^{2}} \left \{
   \left ( \frac{| \vec{u} |}{c} ~-~ \frac{\vec{v} \cdot \vec{S}}{c} \right  ) \vec{S} -
   \frac{\vec{v}}{c} + x' ~\frac{2}{3} ~\frac{R}{c} ~
   \vec{\omega} \times \vec{S} \right \} ~.
\end{equation}

Equation of orbital motion represented by Eq. (35) depends on
angular rotational velocity of the particle.

\section{Conclusion}
The action of solar wind on motion of arbitrarily shaped dust particle
would produce planar motion for non-rotating meteoroid orbiting the Sun.
If angular velocity of rotation of the meteoroid is large enough, then
rotational term may change orbital plane (see Eq. 35 for spherical particle;
see also Eq. 8).  Euler's dynamical equations show that orbital axis of the
particle changes and no equilibrium for the orientation of the axis exists.

Equation of motion and quantities for Euler's dynamical equations for
arbitrarily shaped dust grain are given by Eqs. (2)-(4), (8)-(10), (13)-(14),
(17) and the first part of Eq. (18).

\section*{Acknowledgement}
This work was supported by the Scientific Grant Agency VEGA, Slovak Republic,
Grant No. 2/0016/09.

\end{document}